\documentstyle[12pt,epsfig,amss]{article}

\catcode`\@=11
\long\def\@makefntext#1{
\protect\noindent \hbox to 3.2pt {\hskip-.9pt  
$^{{\ninerm\@thefnmark}}$\hfil}#1\hfill}                

\def\@makefnmark{\hbox to 0pt{$^{\@thefnmark}$\hss}}  
        
\def\ps@myheadings{\let\@mkboth\@gobbletwo
\def\@oddhead{\hbox{}
\rightmark\hfil\ninerm\thepage}   
\def\@oddfoot{}\def\@evenhead{\ninerm\thepage\hfil
\leftmark\hbox{}}\def\@evenfoot{}
\def\sectionmark##1{}\def\subsectionmark##1{}}

\renewcommand{\thefootnote}{\fnsymbol{footnote}}

\newcounter{sectionc}\newcounter{subsectionc}\newcounter{subsubsectionc}
\renewcommand{\section}[1] {\vspace*{0.6cm}\addtocounter{sectionc}{1} 
\setcounter{subsectionc}{0}\setcounter{subsubsectionc}{0}\noindent 
        {\normalsize\bf\thesectionc. #1}\par\vspace*{0.4cm}}
\renewcommand{\subsection}[1] {\vspace*{0.6cm}\addtocounter{subsectionc}{1} 
        \setcounter{subsubsectionc}{0}\noindent 
        {\normalsize\it\thesectionc.\thesubsectionc. #1}\par\vspace*{0.4cm}}
\renewcommand{\subsubsection}[1]
{\vspace*{0.6cm}\addtocounter{subsubsectionc}{1}
        \noindent {\normalsize\rm\thesectionc.\thesubsectionc.\thesubsubsectionc. 
        #1}\par\vspace*{0.4cm}}
\newcounter{appendixc}
\newcounter{subappendixc}[appendixc]
\newcounter{subsubappendixc}[subappendixc]

\renewcommand{\appendix}[1] {\vspace*{0.6cm}
        \refstepcounter{appendixc}
        \setcounter{figure}{0}
        \setcounter{table}{0}
        \setcounter{equation}{0}
        \renewcommand{\thefigure}{\Alph{appendixc}.\arabic{figure}}
        \renewcommand{\thetable}{\Alph{appendixc}.\arabic{table}}
        \renewcommand{\theappendixc}{\Alph{appendixc}}
        \renewcommand{\theequation}{\Alph{appendixc}.\arabic{equation}}
        \noindent{\bf Appendix \theappendixc #1}\par\vspace*{0.4cm}}

\def\abstracts#1{{
        \centering{\begin{minipage}{12.2truecm}\footnotesize\baselineskip=12pt\noindent
        \centerline{\footnotesize ABSTRACT}\vspace*{0.3cm}
        \parindent=0pt #1
        \end{minipage}}\par}} 

\renewenvironment{thebibliography}[1]
        {\begin{list}{\arabic{enumi}.}
        {\usecounter{enumi}\setlength{\parsep}{0pt}
\setlength{\leftmargin 1.25cm}{\rightmargin 0pt}
         \setlength{\itemsep}{0pt} \settowidth
        {\labelwidth}{#1.}\sloppy}}{\end{list}}

\newcounter{itemlistc}
\newcounter{romanlistc}
\newcounter{alphlistc}
\newcounter{arabiclistc}

\newcommand{\fcaption}[1]{
        \refstepcounter{figure}
        \setbox\@tempboxa = \hbox{\footnotesize Fig.~\thefigure. #1}
        \ifdim \wd\@tempboxa > 6in
           {\begin{center}
        \parbox{6in}{\footnotesize\baselineskip=12pt Fig.~\thefigure. #1}
            \end{center}}
        \else
             {\begin{center}
             {\footnotesize Fig.~\thefigure. #1}
              \end{center}}
        \fi}

\newcommand{\tcaption}[1]{
        \refstepcounter{table}
        \setbox\@tempboxa = \hbox{\footnotesize Table~\thetable. #1}
        \ifdim \wd\@tempboxa > 6in
           {\begin{center}
        \parbox{6in}{\footnotesize\baselineskip=12pt Table~\thetable. #1}
            \end{center}}
        \else
             {\begin{center}
             {\footnotesize Table~\thetable. #1}
              \end{center}}
        \fi}

\def\@citex[#1]#2{\if@filesw\immediate\write\@auxout
        {\string\citation{#2}}\fi
\def\@citea{}\@cite{\@for\@citeb:=#2\do
        {\@citea\def\@citea{,}\@ifundefined
        {b@\@citeb}{{\bf ?}\@warning
        {Citation `\@citeb' on page \thepage \space undefined}}
        {\csname b@\@citeb\endcsname}}}{#1}}

\newif\if@cghi
\def\cite{\@cghitrue\@ifnextchar [{\@tempswatrue
        \@citex}{\@tempswafalse\@citex[]}}
\def\citelow{\@cghifalse\@ifnextchar [{\@tempswatrue
        \@citex}{\@tempswafalse\@citex[]}}
\def\@cite#1#2{{$\null^{#1}$\if@tempswa\typeout
        {IJCGA warning: optional citation argument 
        ignored: `#2'} \fi}}

 1
 1
 1

\font\ninerm=cmr9

\newcommand{\xb}{\mbox{$x~$}}  
\newcommand{\xbx}{\mbox{$x$}}  

\newcommand{\Qsq}{\mbox{$Q^2~$}}
\newcommand{\Qsqx}{\mbox{$Q^2$}}

\newcommand{\ftwo}{\mbox{$F_2~$}}

\newcommand{\lambdams}{\mbox{$\Lambda_{\rm \overline{MS}}~$}}

\newcommand{\GeV}{\mbox{\rm ~GeV~}}
\newcommand{\GeVx}{\rm GeV}
\newcommand{\MeV}{\mbox{\rm ~MeV~}}
\newcommand{\GeVsq}{\mbox{${\rm ~GeV}^2~$}}
\newcommand{\GeVsqx}{\mbox{${\rm ~GeV}^2$}}

\newcommand{\ep}{\mbox{$ep~$}}
\newcommand{\alsmz}{$\alpha_s (M_{Z})$}
\newcommand{\als}{$\alpha_s$}

\newcommand{\ee}{\mbox{$e^+e^-$}}
\newcommand{\pp}{\mbox{$p\bar{p}$}}

\begin{document}
%
%
%
%
%
%
\noindent
\vspace{-0.2cm} 
{\tt MPI-PhE/97-22} \hfill {\tt hep-ph/9709240} \\

\centerline{\normalsize\bf 
Jet Production in Deep-inelastic Scattering at HERA}

\centerline{\footnotesize Tancredi Carli}
\baselineskip=13pt
\centerline{\footnotesize\it 
Max-Planck-Institut f\"ur Physik,
Werner-Heisenberg-Institut}
\baselineskip=12pt
\centerline{\footnotesize\it 
F\"ohringer Ring 6, 
D-80805 M\"unchen,
Germany,  
}
\centerline{\footnotesize E-mail: h01rtc@rec06.desy.de}
\vspace*{0.3cm}
\baselineskip=13pt
\centerline{\footnotesize\it 
Invited talk given at the workshop 'New Trends in HERA Physics',}
\centerline{\footnotesize\it Schloss Ringberg, Tegernsee (Germany), May 1997. }
\centerline{\footnotesize\it On behalf of the H1 and ZEUS collaborations.}

\vspace*{0.9cm}
\abstracts{
The main experimental results on jet production
at HERA are reviewed.
A study of jet shapes shows that 
the internal structure of jets is well understood.
The potential to accurately determine the strong coupling constant
using jet rates is discussed.
First attempts to extract the gluon
density in next-to-leading order are presented. 
Although fixed order perturbative QCD is able to describe
the shape of most variables associated to the hard
subprocess leading to jets, it
does not reproduce the absolute dijet rate over 
a large phase space region.
}
 
\normalsize\baselineskip=15pt
\setcounter{footnote}{0}
\renewcommand{\thefootnote}{\alph{footnote}}
\section{Introduction}

HERA colliding $27.5$\GeV positrons on $820$\GeV proton
offers  an ideal testing ground for perturbative
QCD in deep-inelastic scattering (DIS). 

Hard processes can be investigated as function 
of a variable scale, such that the transition from one  
phase space region to another or the interplay between two hard scales,
like e.g. the photon virtuality \Qsq or the transverse energy of a parton,
can be studied. The center of mass energy of 
$\sqrt{s} \approx 300 \GeV$ leads to a  large phase space for hadron
production and to the possibility to observe
clean `jets' in the hadronic final state. 

Jets are event properties relating the unobservable quarks and
gluons to the measurable hadronic final state. 
Quantitative studies of jets require an exact definition
of how to combine the spray of hadrons observed in the detector to jets.
The definition should be suitable for the experimental analysis and
for the theoretical calculation. Moreover, it should lead to
infrared and collinear safe results, which do not change, 
if a low energetic particle is added or if a particle is split
into two.
Jets are usually defined either by cone type algorithms
maximizing the transverse energy flowing through a cone
in the $(\eta,\phi)$ plane\footnote{
The pseudo-rapidity is $\eta=- \ln{\tan{(\theta/2)}}$, 
where $\theta$ is defined with respect to the incident proton.
The proton defines the negative $z$ axis.
$\phi$ is the azimuthal angle.}  or by algorithms 
successively recombining closest particles.
Clustering algorithms iteratively merge pairs of particles
until only a few well separated objects are left. 
The decision to assign a given object to a jet is based
on a distance measure $d_{ij}$ and a resolution parameter
$y_{cut}=d_{ij}/{\rm scale}$. 
Table~\ref{tab:jetalgo} summarizes choices for commonly used clustering
algorithms. Also shown are various procedures for the addition of
particle $4$-momenta
to obtain the jet four-momentum (recombination schemes). 

\begin{table}
\begin{tabular}{|c|c|c|c|}
\hline
Name  &    distance measure $d_{ij}$    & scale   & remnant treatment \\
\hline
JADE  & $ 2 E_i E_j (1-\cos{\theta_{ij}})$ & $W^2$   & $ \sum_h E-P_z$ \\
\cline{1-2}
W     & $ {(p_i + p_j)^2}$ &  &  \\
\hline 
$K_T$ & $2 \min(E_i^2, E_j^2) (1-\cos{\theta_{ij}})$  & $Q^2$ or  &
                                       $d_{ip}=2 \; E_i^2 \; (1-\cos{\theta_i})$ \\
\cline{1-2} \cline{4-4}
long.inv. $K_T$ & 
$2 \min({E_T^2}_i, {E_T^2}_j) (\Delta \eta^2_{ij}+ \Delta \phi^2_{ij}) $ 
            & fixed $E_T^2$      &
                                                     $d_{ip}=2 \; {E_T^2}_i$ \\
\hline
\end{tabular}
\mbox{} \mbox{} \mbox{}
\begin{tabular}{|c|c|c|c|}
\hline
 scheme      &    E or JADE      & P                                  
                                 & E$_0$\\
\hline
combination  &  $p_k = p_i + p_j$ & $E_k = |\vec{p_i} + \vec{p_j}|$      
                                 & $E_k = E_i + E_j$ \\                 
procedure    &                  & $\vec{p_k}=  \vec{p_i} + \vec{p_j}$ 
 & $\vec{p_k}= (E_i + E_j) \, (\vec{p_i}+\vec{p_j})/(|\vec{p_i}+\vec{p_j}|)$\\ 
\hline
\end{tabular}
\vspace{0.1cm}
\fcaption{Summary of the main characteristics of popular jet clustering
algorithms. For the $K_T$ algorithm the remnant is considered as a
particle $p$ with infinite momentum. In the JADE and $W$ algorithm
a pseudo-particle with longitudinal components calculated from the
hadrons in the final state is introduced to account for the remnant.
The combination procedure defines
how the four-momentum $p_i = (E_i,\vec{p}_i)$ 
of two objects $i$ and $j$ are combined to 
an object $k$.
\label{tab:jetalgo}
}
\end{table}

Traditionally cone algorithms have been applied in \pp~collisions
where special care has to be taken to separate particles
belonging to the beam remnant from particles associated with 
the hard subprocess. Cone algorithms are invariant under
longitudinal boosts.
In \ee~collisions, where the initial state is free of color charges
and where every particle can be assigned to jets,
clustering algorithms have been used.
In DIS or \pp~collisions they can also be used
if an additional particle to mimic the hadron remnant is 
put in by hand. 

\begin{figure}
\epsfig{figure=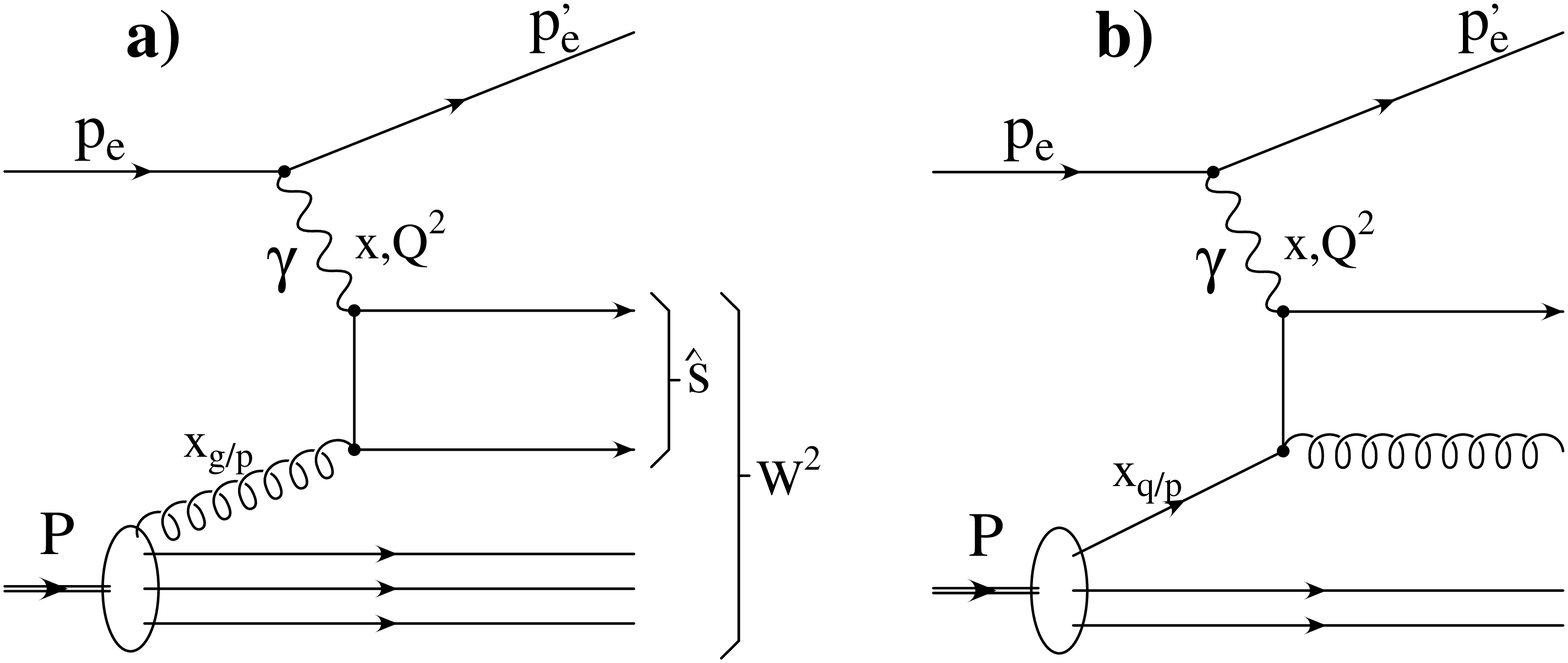,width=10cm}
\begin{tabular}{l}
 \vspace{-5.cm} \\
 DIS variables: \\
   $\Qsq= - q^2 $ \\
   $ x = \Qsq / (2 P \cdot q) $ \\
   $ W^2 = \Qsq (1 - x )/x $ \\
 $\hat{s} = {(j_1 + j_2)}^2$ \\
 $\xi= \xb (Q^2 + \hat{s})/Q^2$ \\
 $z_q= (P \cdot j_1)/(P \cdot q)$ \\
\end{tabular}

\fcaption{Feynman diagrams for the production of $2+1$~jet events 
to first order of \als~in \ep-collisions.
$q$ ($P$) denote the four-momentum of the photon (proton). $j_1$
($j_2$) is the four-momentum of one of the jets associated
to the hard subprocess.
\label{fig:feynjets}}
\vspace{-0.5cm}
\end{figure}

At HERA, events with two jets in the central part of the detector
plus the jet associated with the proton remnant ($2+1$ jet events)
can be produced in a quark ($q\gamma \to q g$) 
or a gluon ($g\gamma \to q \bar{q}$) 
initiated hard subprocess (see Fig.~\ref{fig:feynjets}).
The $2+1$ jet cross section is given by\cite{jet:locross}:
\begin{eqnarray}
\frac{d^2\sigma_{2+1}}{dx dQ^2} \sim \alpha_s(\mu_r)
\int \frac{d\xi}{\xi} 
(
C_{qg      }(x/\xi,z_q,x,Q^2) \cdot q(\xi,\mu_f^2) + 
C_{q\bar{q}}(x/\xi,z_q,x,Q^2) \cdot g(\xi,\mu_f^2) 
)
\label{eq:cross}
\end{eqnarray}
where $\alpha_s$ is the strong coupling constant, 
$\mu_r$  the renormalisation and  $\mu_f$ the factorisation scale.
\xb and \Qsq are the usual variables to inclusively describe 
deep-inelastic scattering (see Fig.~\ref{fig:feynjets} for definition). 
The coefficient functions $C_{q\bar{q}}$ and $C_{qg}$
can be calculated in perturbative QCD.
In addition to \xb and \Qsq they depend\footnote{The dependence of
$C_{qg}$ and $C_{q\bar{q}}$ on the 
azimuthal angle $\phi$ between the lepton and the parton
scattering plane is not given in the formula.} on the 
two variables $\xi$ and $z_q$
characterizing the short distance subprocess which is
responsible for the dijet system.
The fraction of the longitudinal proton
momentum carried by the incoming parton $\xi$,
can be calculated in leading order \als~(LO) from the invariant mass
of the jet system $\sqrt{\hat{s}}$:
\begin{equation}
\xi = x \; ( 1 + \hat{s}/Q^2) 
\vspace{-0.1cm}
\label{eq:xi}
\end{equation}
The variable $z_q$ is given by:
\begin{equation}
z_q = \frac{P \cdot j_1}{P \cdot q} 
\approx \frac{1}{2} (1-\cos{\theta^*}) \approx
\frac{E_j (1 - \cos{\theta_j})}{\sum_{j=1,2} E_j (1 - \cos{\theta_j})}
\label{eq:zq}
\end{equation}
where $q$ ($P$) is the $4$-momentum of the photon (proton) and
$\theta^*$ is the polar angle\footnote{The polar angle is defined
with respect to the incoming parton direction.} 
of one of the jets with $4$-momentum $j_1$
in the photon-parton center of mass frame. $E_j$ and $\theta_j$ are
the energy and polar angle of the jets in the laboratory frame.
The parton distribution functions in the proton 
$q(\xi,\Qsqx)$ and $g(\xi,\Qsqx)$ absorb the collinear and infrared
singularities which occur in the calculations of $C_{qq}$ and $C_{qg}$, 
and they have to be extracted from experiments. 
Once measured at some scale $Q^2_0$, they can be perturbatively
evolved to any other scale, e.g. \Qsqx, 
using the DGLAP equations\cite{th:dglap}. 
By convoluting them with the appropriate coefficient
functions, they can be used to calculate the cross section in 
any other process. In this sense they are universal. 

\begin{figure}
\mbox{\hspace{-2.cm}
\epsfig{width=8cm,file=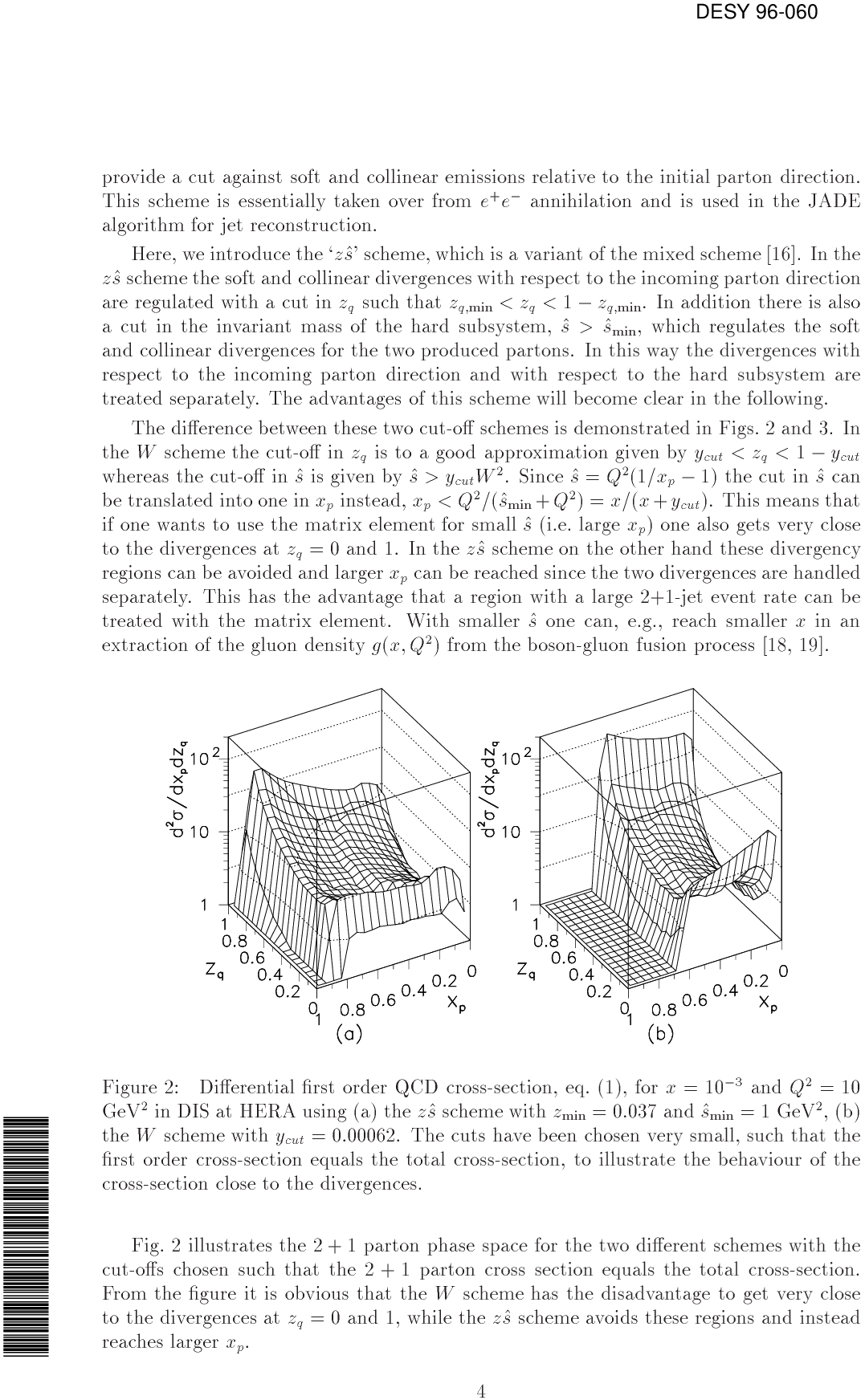,
  bbllx=71pt,bblly=202pt,bburx=296pt,bbury=429,clip=}
}
\mbox{\hspace{-2.cm}
\epsfig{width=10cm,file=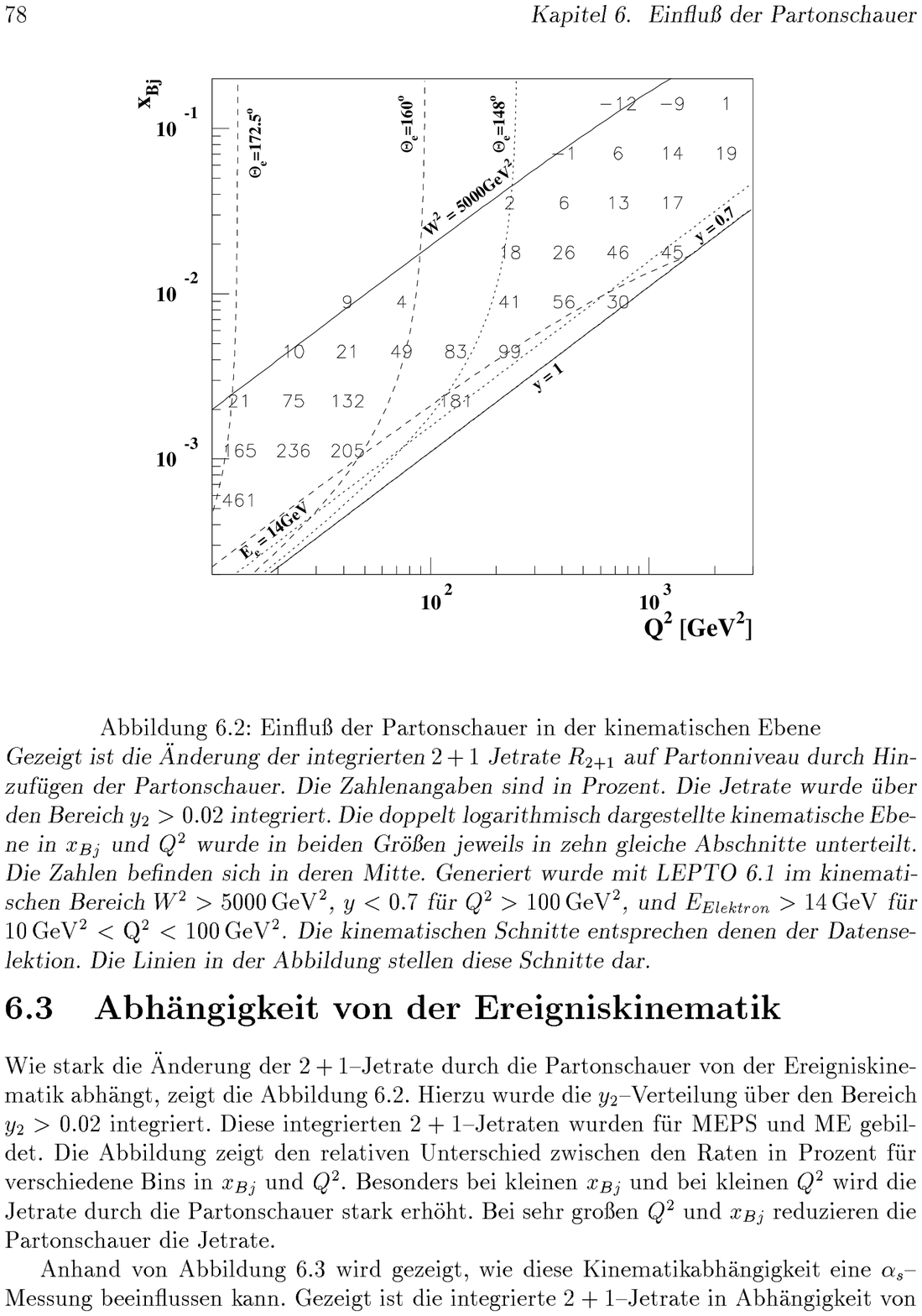,
  bbllx=55pt,bblly=431pt,bburx=475pt,bbury=739,clip=}
}
 \put(-120.,6.) {(b)}

\fcaption{
 a) Differential cross section $d^2\sigma /dx_p dz_q $ 
    in first order of \als~for $x=10^{-3}$ and $Q^2=10$~\GeVsqx.
    The figure is taken from ref.\cite{mc:ingel96}.
 b) Relative change in percent of the dijet rates when adding
    to the leading order cross section higher orders based on
    a leading logarithm parton shower model in the LEPTO Monte Carlo.
    The lines represent typical selection cuts on the scattered
    electron. In addition, a
    cut on $z_q > 0.1$ has been applied. 
    The jet rate is based on the JADE algorithm for $y_{cut}=0.02$.
    The figure is taken from ref.\cite{jet:flammthese}.    
 }
\label{fig:mefig1}
\end{figure}

In leading order of \als,
the coefficient functions diverge like $C_{q\bar{q}} \sim 1/( z_q (1-z_q))$ and
$C_{qg} \sim 1/( (1-x/\xi) (1-z_q))$ 
and have to be regulated by a cut-off. 
In Fig.~\ref{fig:mefig1} is shown how the cross section
$d^2\sigma/dx_p dz_q$ (where  $x_p=\xb/\xi$) increases towards low $z_q$
and $x_p$ and towards $z_q = 1$.  
It is clear that in the regions where the LO cross section diverges,
large contributions from higher orders can be expected.

The dijet cross section including
higher order parton emissions are difficult to calculate using exact
expressions for the coefficient functions. 
Since the phase space integrals cannot be solved analytically,
numerical methods have to be applied.
Recently several Monte Carlo integration programs
computing the cross section to next-to-leading order (NLO) became
available. 
Early attempts like PROJET\cite{jet:projet} or
DISJET\cite{jet:disjet} used a semi-analytical approach 
which only allowed jets to be defined by the $W$ algorithm 
with the JADE scheme\cite{jet:jade} and were
restricted to a small set of calculable distributions. Moreover,
in these programs approximations were made which turned out not to be valid
over the full phase space\cite{jet:mepjet}. 
The programs MEPJET\cite{jet:mepjet}, based on a phase
space slicing method\cite{jet:phasepaceslicing}, 
and DISENT\cite{jet:disent}, based on a method to subtract
the divergences from individual contributions\cite{jet:subtraction} and on
an analytical treatment of specific divergent terms\cite{jet:dipolfact},
are flexible NLO Monte Carlo programs which allow
any jet definition scheme and
arbitrary experimental cuts to be analyzed.
For $\Qsq > 40 \GeVsq$ and for jets defined by the $k_T$ 
algorithm\cite{jet:kt}  the agreement of the $2+1$ and $3+1$ jet cross 
sections between the two programs is found to be 
on the $3\%$ level\cite{jet:mirkesdis97}.

Higher order parton emissions and models for the transition of partons
into hadrons are only included in Monte Carlo programs
incorporating coefficient functions to leading order  such as
ARIADNE\cite{mc:ariadne}, LEPTO\cite{mc:lepto}, HERWIG\cite{mc:herwig},
PYTHIA\cite{mc:pythia} and RAPGAP\cite{mc:rapgap}. 
They model multi-parton emission
by either parton showers\cite{mc:partonshower} (PS) based on the
DGLAP equations or by a chain of 
independently radiating colour dipoles\cite{mc:dipole} (CDM).
HERWIG uses the cluster hadronisation model\cite{mc:cluster}. 
In all other programs 
hadronisation is based on the LUND string model\cite{mc:string} 
as implemented in JETSET\cite{mc:jetset}.

\begin{figure}
\epsfig{width=7cm,height=5cm,file=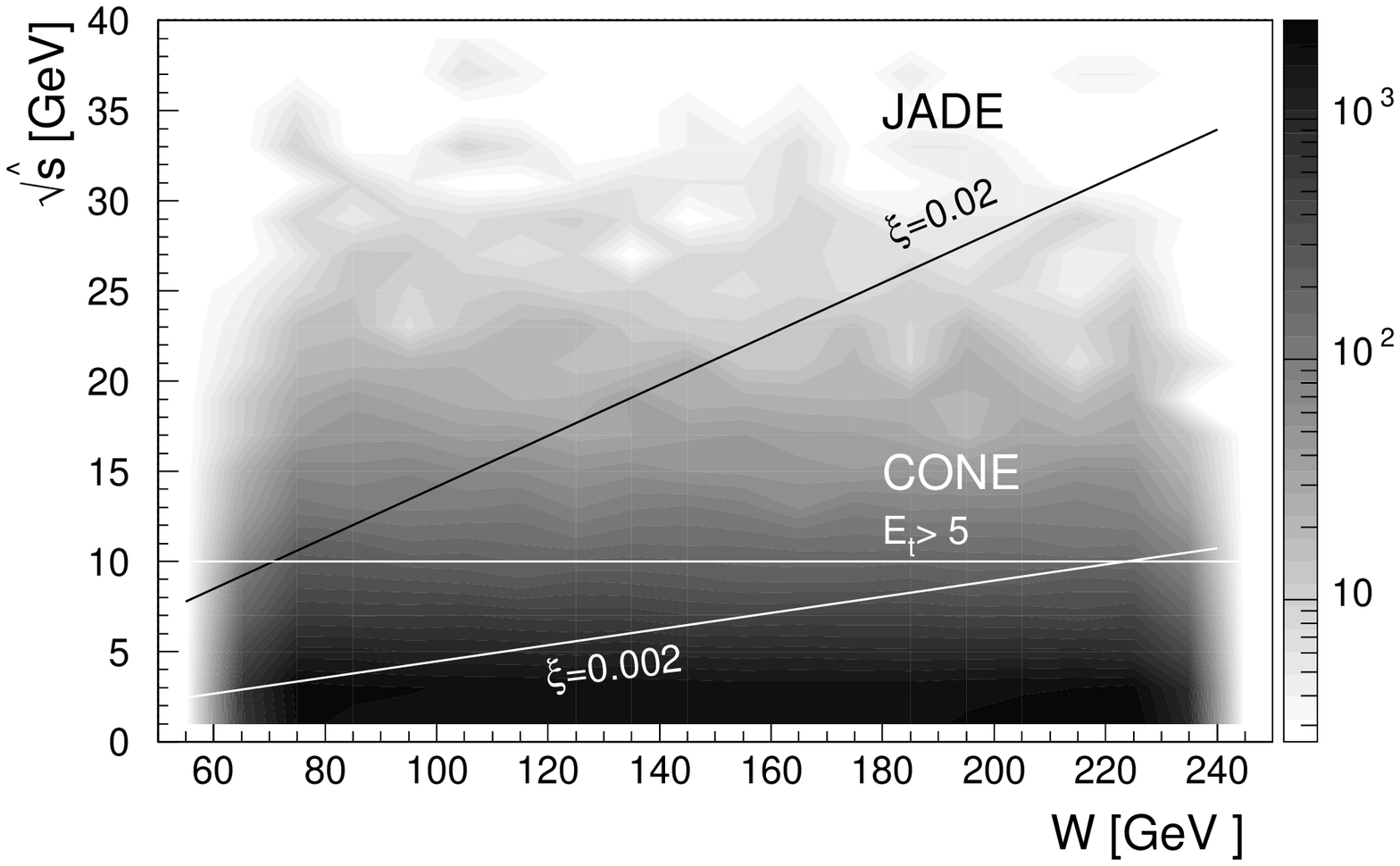}
\epsfig{width=7.5cm,height=5cm,file=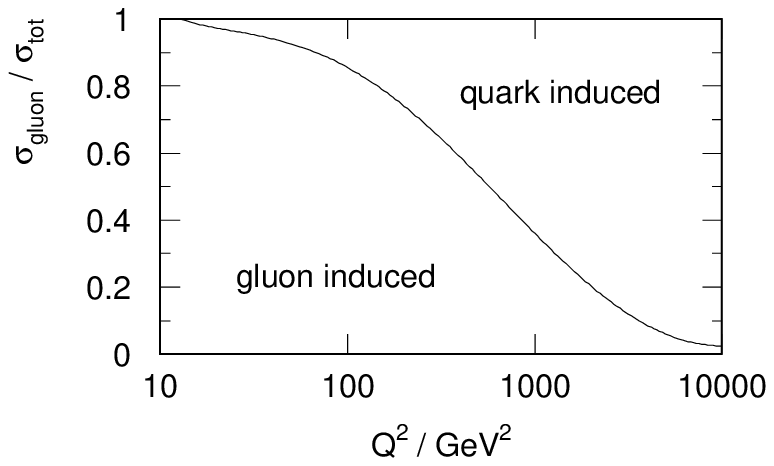}
\fcaption{ a) Rate of $2+1$ jet events as function of the invariant
 mass of the two jet system $\sqrt{\hat{s}}$ and the invariant mass
 of the hadronic final state $W$. Indicated are the phase space restrictions
 introduced when using the cone or the JADE algorithm.
 b) Fraction of gluon and quark induced processes leading 
    to $2+1$ jet events with $E_T > 5$\GeV and  $-1 <\eta < 2$
    defined by a $K_t$ algorithm in the Breit frame
    as function of \Qsqx. The cross section has been calculated in
    NLO in the kinematic range
    $\xb > 1 \cdot 10^{-4}$ and $0.15 < y< 0.65$.
 }
\label{fig:mefig2}
\end{figure}

The relative change in percent of the dijet rates defined 
with the JADE algorithm for $y_{cut}=0.02$ when adding
to the LO cross section higher orders based on
a leading logarithm parton shower model (MEPS)
is shown in Fig.~\ref{fig:mefig1}
in the \xb-\Qsq plane applying typical analysis cuts on the
scattered electron.
While for $\Qsq > 100$~\GeVsqx, the relative change only exceeds
$20\%$ in the low \xb region, parton showers change the dijet rate
by a large factor for $10 < \Qsq < 100$~\GeVsq and
$1 \cdot 10^{-4} < \xb < 2 \cdot 10^{-3}$.
The influence of parton emissions beyond leading order  
does also depend on the jet variables $z_q$ 
and $\xi$. Since different jet algorithms
cover different phase space regions, it is always important 
to specify which jet definition scheme 
was used. Fig.~\ref{fig:mefig2} illustrates the rate of
$2+1$ jet events as function of the invariant mass of the
jet system $\sqrt{\hat{s}}$ and the invariant mass of the whole
hadronic final state $W$. The rate rapidly increases
towards low $\sqrt{\hat{s}}$. Indicated as lines are the phase space
boundaries introduced by the JADE and the cone algorithm.
With a cut on the transverse energy of the jets $E_T > 5$\GeV
the minimum $\sqrt{\hat{s}}$ which can be reached is $10$~\GeVx,
since (in the limit of massless partons) 
from $E_T^2= \hat{s} \, z_q \, ( 1 - z_q )$ follows
$\sqrt{\hat{s}}/2 > E_T$. When using the JADE algorithm
with a resolution parameter 
$d_{ij}/W^2 \approx \hat{s}/W^2 \approx \xi$,
low values of $\sqrt{\hat{s}}$ can only be reached for
low (high) values of $W$ (\xbx). If one takes $y_{cut}=0.02$,
$\sqrt{\hat{s}}$ is below $10$\GeV for $W \lesssim 70$~\GeVx.
The JADE algorithm with a high resolution parameter is therefore
more restrictive than the cone algorithm. In addition to the region
of low $\sqrt{\hat{s}}$, where collinear and infrared divergences
become important, it naturally avoids the region of high $W$ (low \xbx) 
where a large phase space for multi-parton emission is available.
However, the divergences at small and
large $z_q$ can only be avoided by a large $y_{cut}$.

Multi-jet production allows a variety of quantitative tests of our 
understanding of QCD dynamics.
In a phase space region where the data are well described by
perturbative QCD, HERA offers the opportunity to
extract \als~over a wide range of \Qsq in a single experiment.
This can be done at large \xb and \Qsq where 
the parton densities in the proton are well constrained. 
At low \Qsq and correspondingly low values of
$\xi$, gluon initiated processes become increasingly
important. Here, the measured $2+1$ cross section can be used
to determine the gluon density assuming a value of \als.
The sensitivity to both \als~and $g(\xi,\Qsqx)$ could also be used
to simultaneously extract these quantities. Such an analysis is however
theoretically and experimentally difficult.
In the region where the data significantly differ from what
is expected by a fixed order calculation of perturbative QCD,
jet observables are useful tools to  
gain insights in the complex parton dynamics occurring in electron
proton collisions.

\section{Jet shapes in DIS}
\begin{figure}
\vspace*{13pt}
\epsfig{width=7cm,file=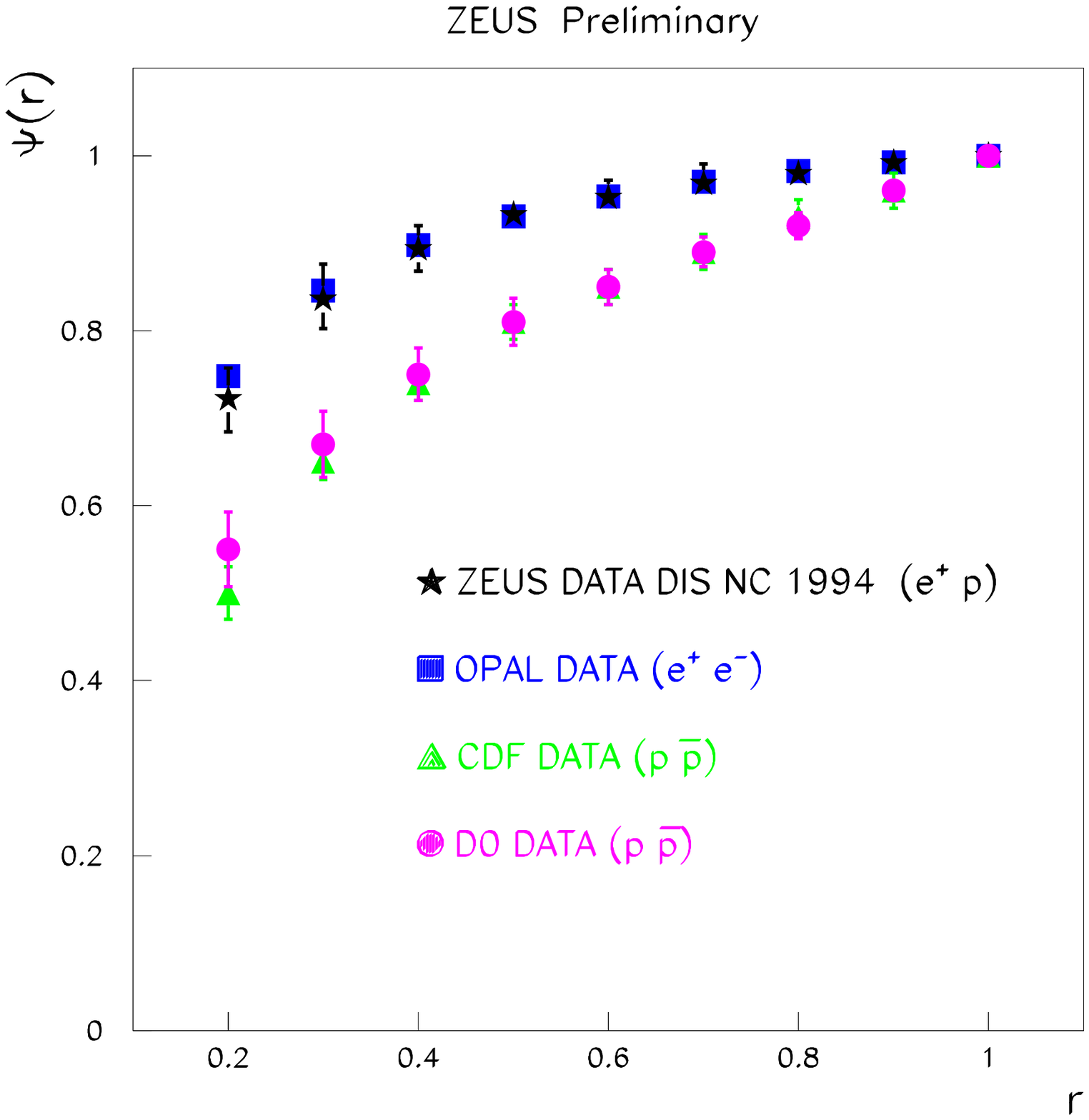}
\epsfig{width=7cm,file=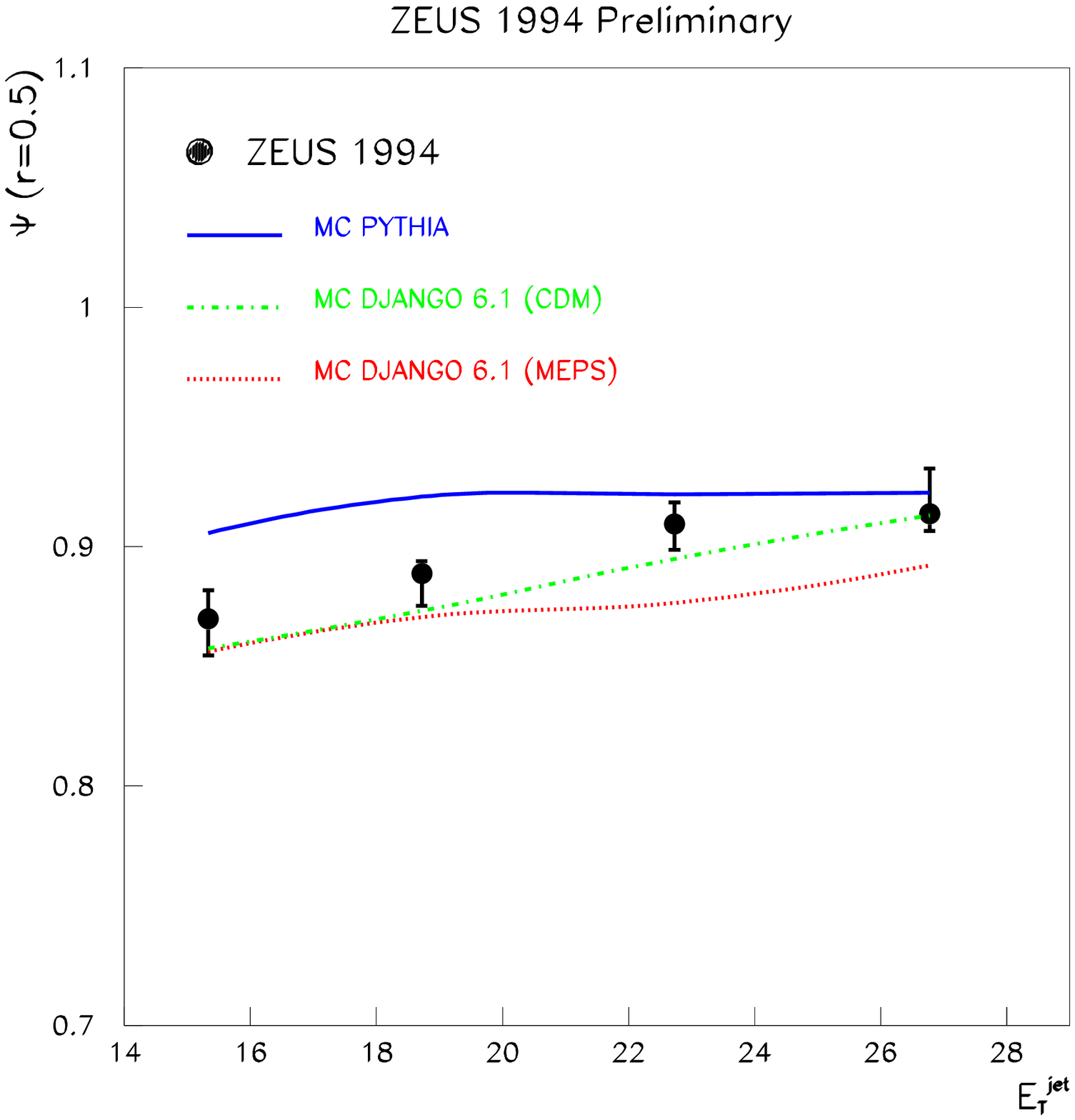}
\fcaption{
a) Fractional transverse energy $\Psi(r)$ in a jet defined by a 
 cone algorithm with radius $R=1$ and $37 < E_t < 45$\GeV in the
 laboratory frame as function of
 $r= \sqrt{\Delta\eta^2+\Delta\phi^2}$. For OPAL jets with $E_T > 35$\GeV
 were selected, for CDF (D0) the considered $E_T$ range of the jets 
 is $40 \, (45) < E_T < 60 \, (70)$~\GeVx.
b) $\Psi(r=0.5)$ as function of the transverse energy of the jet.
QCD Monte Carlo models are superimposed. They either
calculate the dijet cross section in LO and add 
multi-gluon emissions based on the colour dipole
model (CDM) or by parton showers (MEPS) 
or treat QCD effects only by parton showers (PYTHIA).
}
\label{fig:jetshapes}
\end{figure}

The internal structure of jets provides
useful information on the transition of a parton to the
complex aggregate of hadrons which can be observed in the
detector. The capability to describe detailed properties of jets
is also an important ingredient when comparison of perturbative
calculations with experimental data.

To study the internal structure of jets
in the sample of neutral current DIS events
with $\Qsq > 100$\GeVsq
jets are found in the laboratory system
using the CDF-cone algorithm\cite{jet:cdfcone}
with a cone radius $R=\sqrt{\Delta\eta^2 + \Delta\phi^2}=1$.
Only jets are considered which have in the laboratory frame
a transverse energy $E_T > 14$\GeV and lie in the rapidity
range $-1 < \eta < 2$.
The jet shape is defined as the average fraction of the transverse 
energy of the jets inside an inner cone $r$ 
concentric to the outer jet cone with radius $R$:
\begin{equation}
\Psi(r) = \frac{1}{N_{\rm jet}} \sum_{\rm jets} 
          \frac{E_T(r) }{E_T(R) } \; \; , \; \;
           E_T(r)=\int_0^r dr' dE_T(r')/dr' 
\end{equation}
where $E_T(r)$ is the transverse energy
within an inner cone of radius $r$ and $N_{\rm jet}$ is the total
number of selected jets. By definition $ 0< \Psi < 1$ and
$\Psi(R)=1$. The steepness of the rise of $\Psi$ describes the
collimation of the jet. 

The jet shape $\Psi$ corrected to hadron level for jets
with $E_T>37$\GeV as function of $r$ 
is shown in Fig.~\ref{fig:jetshapes}a. $90$\% of the
jet transverse energy is already contained in a cone with
$r=0.4$ around the jet axis. This is very similar to jets
produced in \ee~collisions\cite{jet:opaljetshapes} 
with $E_T > 35$~\GeVx. The jets selected from \pp~collisions
for $40 \, (45) < E_T < 60 \, (70)$~\GeV measured by 
CDF (D0)\cite{jet:cdfjetshapes,jet:d0jetshapes}
are considerably broader. This might be explained by
the fact that in \ee~and \ep~collisions the partons initiating
the cascade leading to the jet
are mainly quarks while in \pp~collisions they
are gluons.

The jet shape depends on the transverse energy
of the jet. Jets get narrower with increasing $E_T$ as is shown
in Fig.~\ref{fig:jetshapes}b. 


Monte Carlo models based on a hard subprocess in LO plus 
additional multi-gluon emissions are able to describe the data.
Also PYTHIA where QCD effects are only treated by parton showers
reproduces the main features of the data.
The best description of the $E_T$ dependence of $\Psi$ is
obtained when partons are emitted according to the colour
dipole model. 
The internal jet structure in DIS at HERA is fairly well reproduced
by all QCD models.

\section{Determination of the strong coupling constant}

The $2+1$ jet rate defined with the JADE algorithm
depends on the jet resolution parameter $y_{cut}$. 
For each event $y_2$ is defined as the value of $y_{cut}$ 
where a $1+1$ jet event switches to a $2+1$ jet event.
\begin{figure}
\vspace{-1.cm}
\epsfig{width=12cm,file=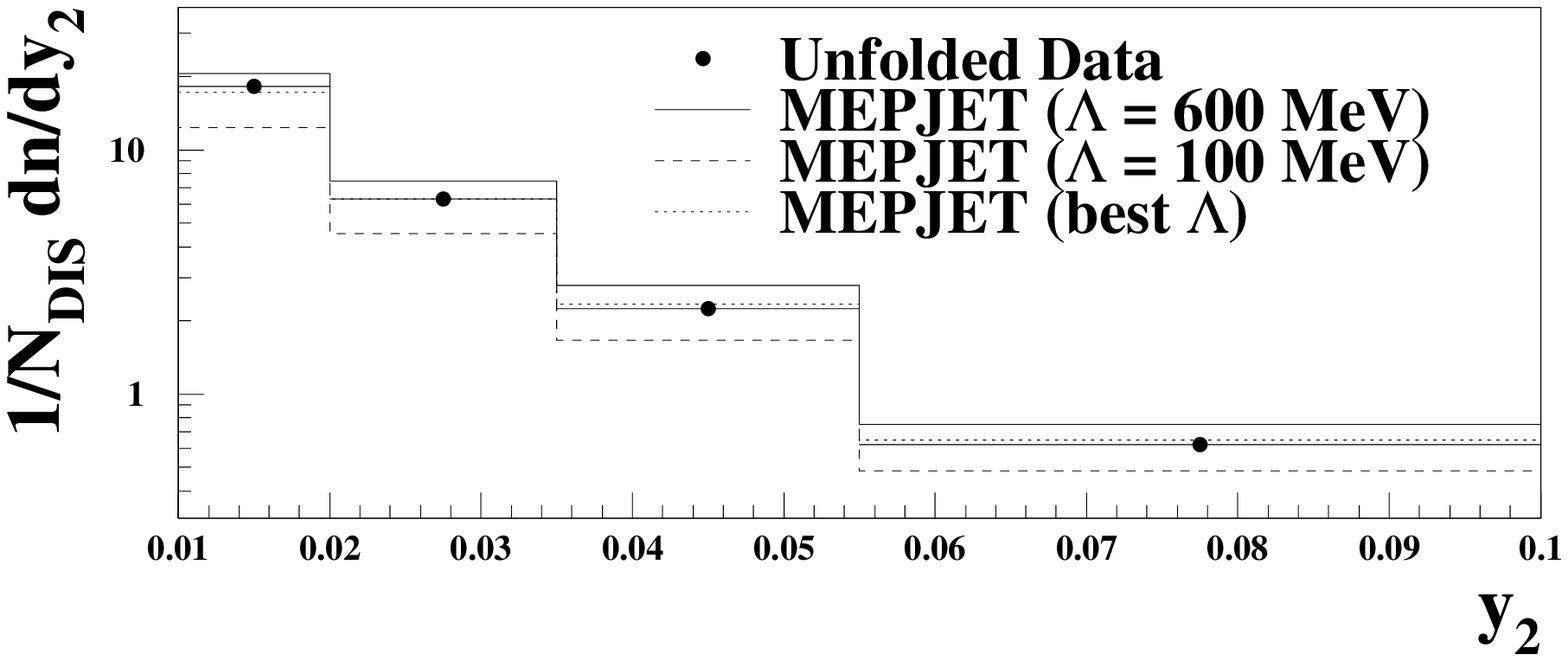}
  \put(-200.,150.) {H1 preliminary }
\fcaption{Distribution of the resolution parameter $y_2$ in the JADE
 algorithm where a $1+1$ jet event switches to a $2+1$ jet event.
 Shown are unfolded data and the next-to-leading order
 prediction for some values of $\Lambda^{(4)}_{QCD}$. 
 $N_{\rm DIS}$ is the total number of DIS events in the 
 considered kinematic region.
}
\label{fig:y2}
\end{figure}

A region of phase space is selected where NLO QCD
can provide a reasonable description of the data, and where effects 
from multi-gluon emissions are small. In the H1 analysis\cite{jet:weberdis97}
this region is defined by:
$Q^2 > 200 \GeVsq$ and $W^2 > 5000 \GeVsqx$.
To avoid events where the invariant mass of the two jet
system is small and where the cross-section diverges,
$y_2$ is restricted to $y_2 > 0.01$.
The $y_2$ distribution, corrected to parton level by an
unfolding method\cite{jet:blobelunfold}, is shown in
Fig.~\ref{fig:y2}. Overlayed is the prediction of NLO
QCD  for three values of \als~(respectively $\Lambda_{QCD}$).
In all bins a strong sensitivity to \als~is observed.
For one given value of \als~NLO QCD gives an excellent
description of the data. The precision which can be expected
from this measurement is of the order of 
$\Delta \alpha_s(M_Z)/\alpha_s(M_Z) \approx 5 - 10$\%.

Keeping $y_{cut}=0.02$ fixed, the number of
$2+1$ jet events ($N_{2+1}$) normalized to the sum of $1+1$ and $2+1$ 
jet events ($N_{2+1} + N_{1+1}$) can be measured and is sensitive to \als. 
Only jets in the acceptance region of  
$10^{o} < \theta_{jet} < 145^{o}$ and with $z_q >0.1$
are counted.
The fractional jet rate
$R_2= N_{2+1}/(N_{1+1} + N_{2+1})$ can be expressed by:
\begin{equation}
R_2(Q^2,y_{cut})= A(Q^2,y_{cut}) \, \alpha_s(Q^2) +
                  B(Q^2,y_{cut}) \, \alpha_s^2(Q^2)
\end{equation}
From the coefficients $A$ and $B$ which can be calculated in NLO
QCD 
and the measured $R_2$, \als$(\Qsqx)$
can be extracted in different \Qsq bins.  
Using the renormalisation group equation these
independent measurements can be used to fit \alsmz.
\begin{figure}
\vspace*{13pt}
\begin{center}
 \mbox{\hspace{-1.5cm}
 \epsfig{figure=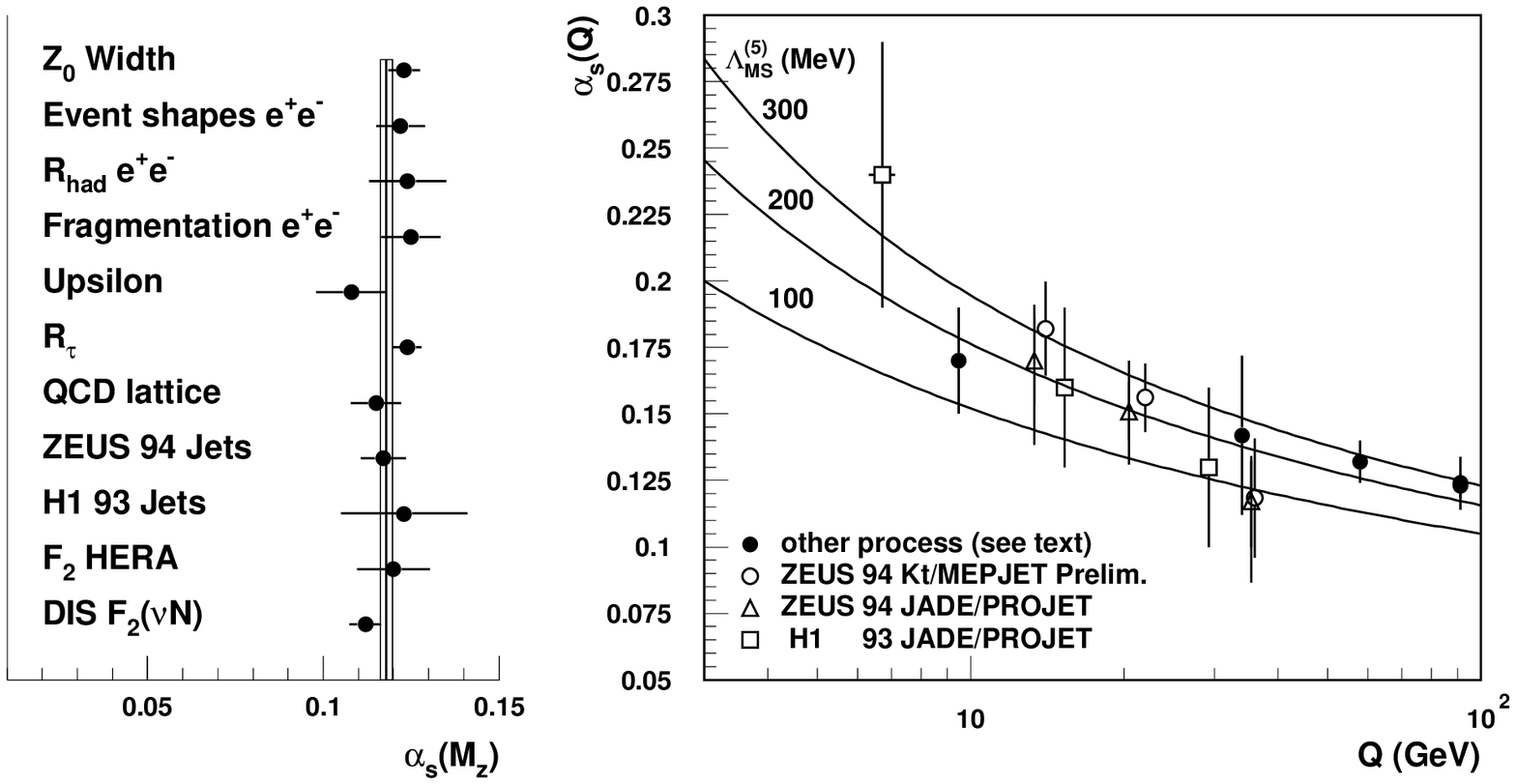,width=17.cm}
}
\end{center}
\vspace{-0.2cm} 
\fcaption{Left: Values and total error of $\alpha_s(M_Z)$ from 
 various processes. 
 The solid line indicates the world average 
 and the band its total error. 
 Right: \als($Q$) from HERA (open symbols) and other processes
 with increasing $Q$ (closed circles): 
 $\Gamma_{\! \Upsilon}$ and
 $\sigma_{\rm \! had}/\sigma_{\rm \! tot}$,
 event shapes and $\Gamma_{\rm \! hadron}/\Gamma_{\rm \! lepton}$ 
 in \ee. The preliminary points using the $k_T$ algorithm
 include the statistical errors only.
 The predictions of the renormalisation group
 equation for three values of \lambdams~are superimposed as lines. 
\label{fig:alphas}}
\end{figure}

For this measurement only the kinematic range of high \Qsq 
and high \xb is considered in order to reduce the uncertainties 
stemming from higher order gluon emissions and from the hadronisation of a complex
partonic configuration. The theoretically and
experimentally difficult region towards the proton remnant is excluded.
Three values of \als($Q^2$) extracted
by the H1\cite{jet:h1alphas} and ZEUS\cite{jet:zeusalphas}
collaboration 
are shown in Fig.~\ref{fig:alphas}. For increasing values of
$Q^2$, \als($Q^2$) decreases as predicted by the renormalisation
group equation shown for \lambdams~between $100$ and
$300$\MeV~(solid lines).
An extrapolation to \als($M_z$) yields:
\begin{eqnarray}
\vspace{-0.3cm}
\label{eq:alphas}
 \nonumber \! \! \! \! \! \! \! \! \! \! 
 {\rm H1 \; 93:  } \; \; \; \; \; \; \; \; \; \; 
  \alpha_s(M_z)  = 0.123 \pm 0.012 \, {\rm(stat)}
 \pm 0.013 \, {\rm (syst.)} \\
 \nonumber
 {\rm ZEUS \; 94:} \; \; \alpha_s(M_z) = 0.117
 \pm 0.005 \, {\rm (stat)} \pm 0.005 \, {\rm(exp.)} \pm 0.007 \, {\rm (th.)}
\end{eqnarray}
which is consistent with other values obtained from
a large variety of different processes as shown in Fig.~\ref{fig:alphas}
(for ref. see\cite{jet:pdg}).
The agreement found in \ee, \ep~and \pp~collision for reactions
at very different scales is an important and successful test of QCD.
The error on the HERA measurements are even with the present
limited statistics already competitive.
The dominant experimental errors stem from the uncertainty on the
energy scale for hadrons, the model dependence of assigning
partons to jets and the phenomenological description of
the hadronisation process. Furthermore the renormalisation scale
and the choice of the input parton density systematically influences
the result. 
These results have been obtained with the NLO program PROJET or DISJET.

When particles are combined to jets, the details of how the $4-$momenta
are added are important, 
in particular, because higher order QCD calculations are only carried
out for massless partons whereas the jets acquire a mass during the
combination procedure. 
In the E-scheme $4$-momenta are added to
massive jets while in the $E_0$ and $P$-schemes jets remain massless
by rescaling either the $3$-momenta or the energy (see table~\ref{tab:jetalgo}
for details).
The dependence of the jet cross section on the recombination 
schemes (RS) only appears in the calculation when jets are composed
out of more than one parton. 
Corrections of $\cal{O}$($\alpha_s^3$) due to the different 
recombination schemes are potentially large.
When applying different recombination schemes to the data,
the measured 2+1 jet cross sections defined with the $E$ and $E_0$ 
algorithms agree within few percent, but are lowered by up to $20\%$
for the $P$-scheme\cite{jet:rosenbdis96}.
In PROJET only the E scheme is implemented.
It is therefore not surprising that the extracted \alsmz~agrees  
within $3$\% when the E and E0 scheme is only applied to the data, 
but gives an inconsistent result 
when using the P scheme\cite{jet:rosenbdis96}. 
An additional problem is that in the NLO calculation the $W$ algorithm
is implemented, while in the experimental analysis the JADE algorithm
has to be used, i.e. mass terms in $d_{ij}$ are neglected.


The new NLO programs MEPJET and DISENT 
reveal a strong dependence on the 
recombination scheme\cite{jet:mirkesdis96} when using the JADE or $W$
algorithm. In addition lead these algorithms
to large NLO corrections of the dijet cross section.
%

The cone or $k_T$ jet algorithms seem to
be better suited for precision QCD tests,
since the variation of the cross section when going from leading
to next-to-leading order is much smaller\cite{jet:mirkeskrakow}. 
Moreover, these jet
algorithms are less dependent on the choice of the renormalisation
and factorisation scale
and allow a larger phase space to be covered.

 The ZEUS collaboration has reanalyzed their data\cite{jet:trefzger} 
 using the $k_T$ algorithm. The preliminary values of \als($Q$) obtained
 in three bins of $Q$ are shown\footnote{Only the statistical
 errors are included in the figure.} in Fig.~\ref{fig:alphas}.
 They are consistent with the results obtained
 with the JADE algorithm. The value extrapolated to the $Z^0$ boson mass
 is: $\alpha_s(M_Z)= 0.118 \pm 0.008 \, ({\rm stat})$. 

\section{Determination of the gluon density}

 The parton distributions are well constrained for larger \xb where a lot
 of data are available. However, for low \xbx, 
 where gluons play a crucial r\^ole, they are not well known.
 In this regime, HERA has the unique possibility to investigate 
 the distribution of the gluon density in the proton.
 The scaling violation of the proton structure function 
 \ftwo~offers an indirect way to get a handle on the gluon density.
 This measurement\cite{h1:f2of94,z:f2gluon} 
 is possible because of the extremely successful
 description of \ftwo~by the NLO DGLAP equations.
 Observables based on $2+1$ jet event can - by including them in
 the global analysis - provide a further constraint on the gluon
 density. Dijet rates can be calculated in 
 next-to-leading order and are sensitive to the gluon density,
 since for $10 \lesssim \Qsq \lesssim 100$ \GeVsq the $2+1$ jet cross
 section is dominated by gluon initiated processes. In this
 kinematic region less than about $20$\% of the $2+1$ jet events 
 are induced by quarks (see Fig.~\ref{fig:mefig2}).

 An alternative approach is to only base the analysis on
 the $2+1$ jet rate. 
 Since large invariant jets masses ($\hat{s}>100$\GeVsqx) 
 are experimentally required to define clean jets, the
 very low \xb regime can not be reached. These measurements are
 nevertheless important to fill the gap between the results 
 from small-\xb scaling violations in \ftwo~and 
 the data on direct photon and jet production in \pp~collisions
 reaching to large-\xbx\cite{jet:martingluon}. 
\begin{figure}
\vspace*{13pt}
%
\epsfig{width=7cm,file=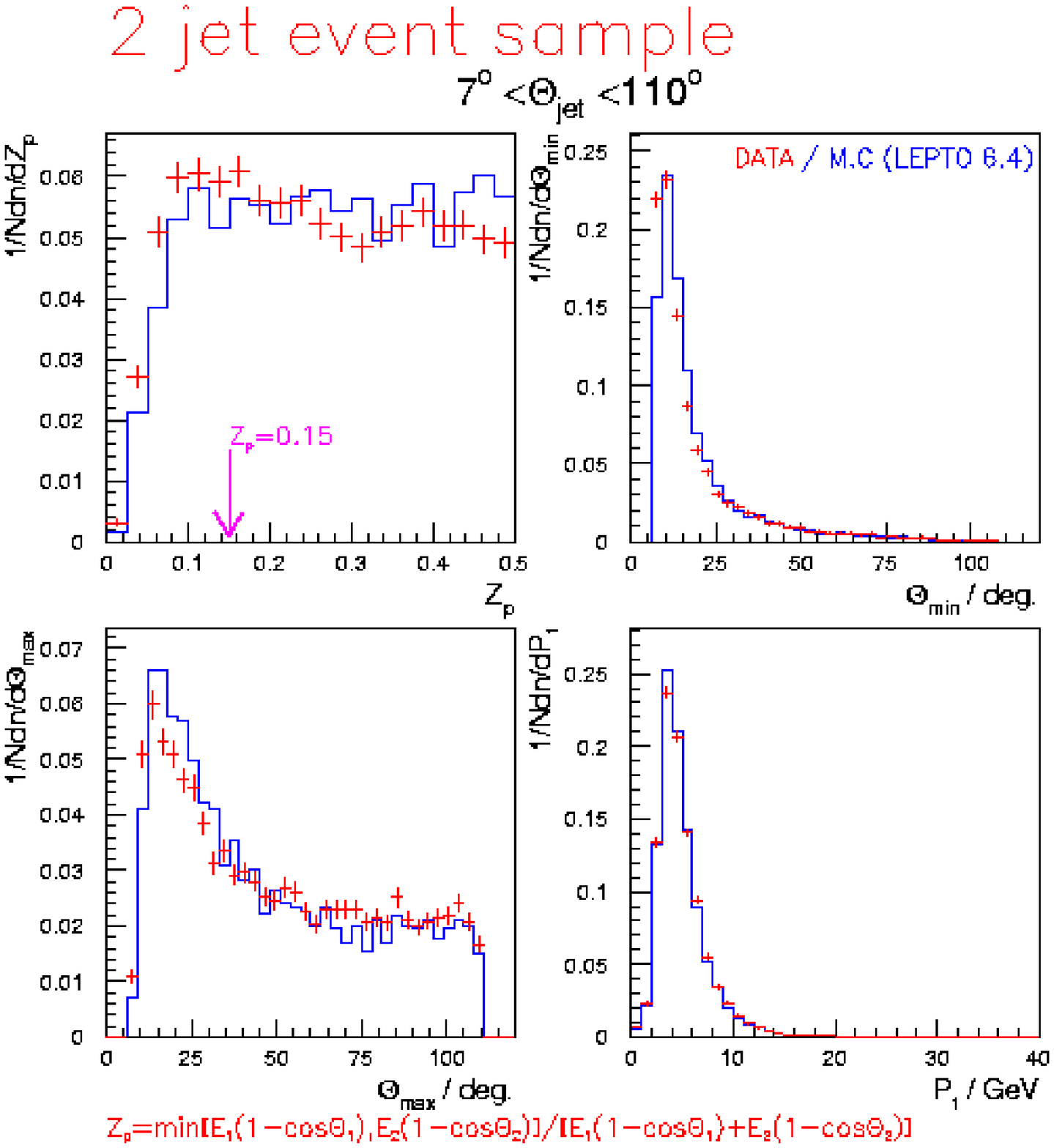,
 bbllx=273pt,bblly=163pt,bburx=532pt,bbury=398,clip=}
\epsfig{width=7cm,file=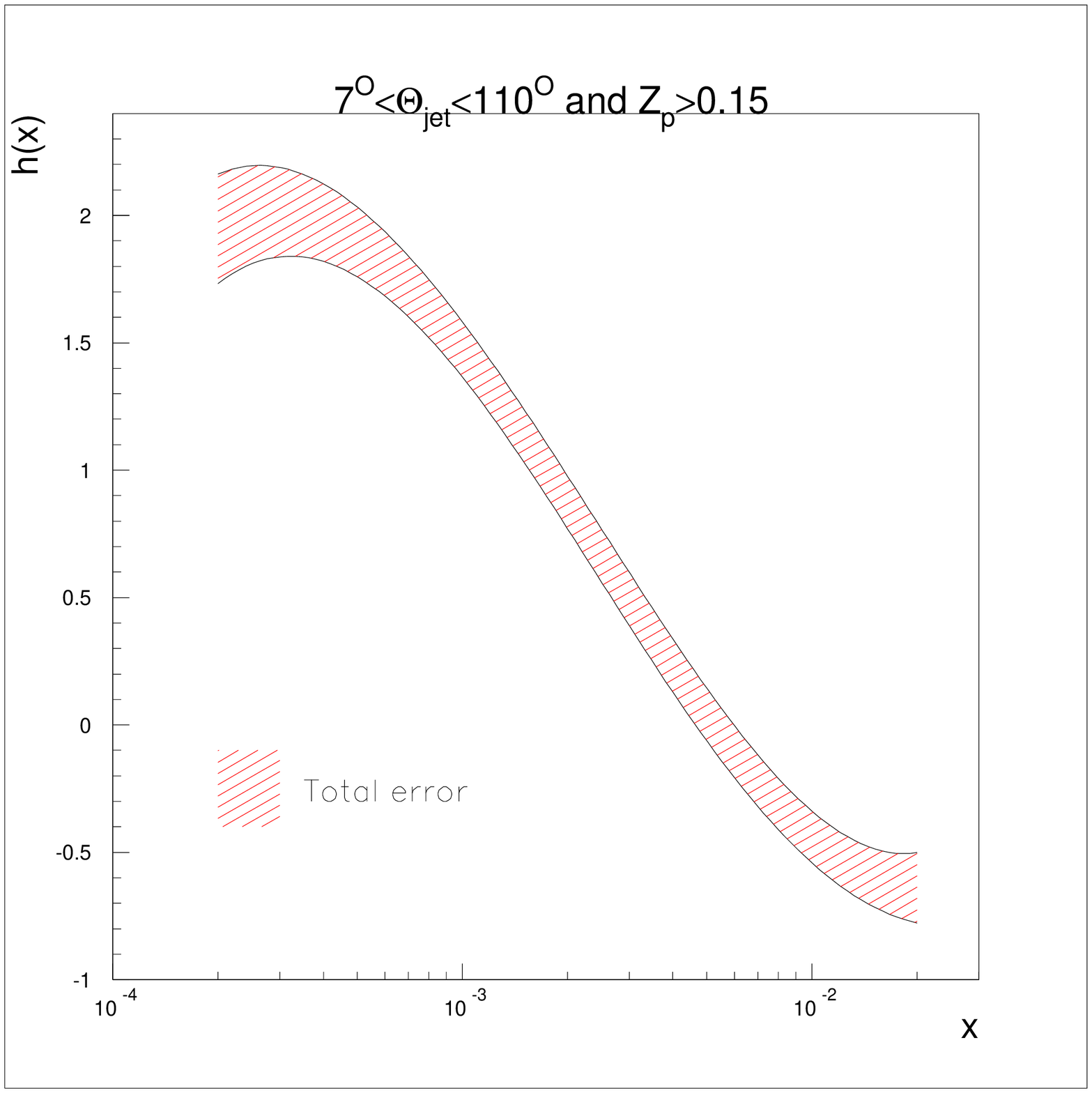,
 bbllx=25pt,bblly=164pt,bburx=511pt,bbury=668,clip=}
 \put(-300.,150.) {H1 preliminary }
 \put(-80.,150.) {H1 preliminary }
\fcaption{
Left: Shape normalised distribution of the transverse energy of jets 
   defined with the
   $k_t$ algorithm (in the Breit frame) using $Q^2/2$ as resolution parameter.
   The data have not been corrected for detector effects.
   Superimposed is a simulation based on LEPTO.
Right: The function $h(\xbx)$ describing the non-perturbative contribution
   of the dijet cross section as defined in the H1 analysis to extract
  the gluon density in NLO.}
\label{fig:gluon}
\end{figure}

 The H1 collaboration\cite{jet:h1logluon} has already exploited
 the $2+1$ jet production to extract the gluon density in LO in the range
 $2 \cdot 10^{-3} \lesssim \xi \lesssim 0.18$.
 When processes where only one gluon is involved are considered, $\xi$ is
 directly observable and the determination of $g(\xi,\Qsqx)$
 is straight forward. This is however only of limited theoretical interest,
 since the LO dijet cross section is subject to large higher
 order corrections. There is moreover a large dependence on the
 arbitrary choice of the renormalisation and factorisation scale. 
 Only, in NLO a reliable prediction of the dijet cross section
 is possible.
 The $2+1$ jet cross section, for instance,
 defined with a cone algorithm for jets with $E_T > 5$\GeV
 remains constant within $5$\% when varying the scale by  
 $3$ orders of magnitude while in LO variations of $40$\% are
 calculated\cite{jet:mirkeskrakow}.

 A NLO extraction of $g(\xi,\Qsqx) $ requires, however, an
 iterative procedure where the input gluon
 density used in the NLO calculation is adjusted until the
 theoretical dijet cross sections agree with the measurement.
 This is most efficiently done by expanding the gluon density
 for a discrete set of points $(\xi_1, ... , \xi_n)$
 using appropriate test functions
 $g(\xi,\Qsqx) =  \sum_{n=1}^{n_{max}} a(\xi_n,\Qsqx) \, T_n(\xi)$
 such that the integral 
 $ \sigma_g = \sum_{n=1}^{n_{max}} a(\xi_n,\Qsqx)
 \int d\xi/\xi \, C_{q\bar{q}}(x/\xi,\Qsqx) \, T_n(\xi)$ 
 in equation \ref{eq:cross}
 must only be calculated once. The coefficient $a$ can then be used
 in the fitting procedure.
 Preliminary results using
 an elegant technique based on Mellin moments\cite{jet:mellingluon},
 where the test functions take the form $T_n= {(x/\xi)}^n$,
 have been presented by the H1 collaboration\cite{jet:rosenbdis96}
 covering the range $0.01 < \xi < 0.1$ using the JADE algorithm.

Another approach also pursued by the H1 collaboration\cite{jet:zomerdis97}
is to use as test functions cubic splines\cite{jet:loboheraws96} with
the property $T_i(\xi_j) = \delta_{ij}$
and to introduce the $2+1$ jet rate as additional
observable in a global fit analysis minimizing 
$\chi^2 = \chi^2(F_2) + \chi^2(R^{det}_{2+1})$.
This has the advantage that no assumption
on the quark densities is needed. The correlation of quarks
and gluons via the DGLAP equation is naturally taken into account.
$R^{det}_{2+1}$ as measured in the detector is for each considered
kinematic bin related to the NLO calculation of the
dijet cross section $\sigma_{2+1}^{\rm pert}$ performed by DISENT:
\begin{equation}
R_{2+1}^{\rm det} = 
\frac{ {\rm MIG}_{2+1}^{1+1} \cdot 
       (\sigma_{2+1}^{\rm pert}(\xbx,\Qsqx) +
        h_{2+1}^{\rm non-pert.}(\xbx,\Qsqx)) 
     }{\sigma_{tot}(\xbx,\Qsqx)}
\end{equation}
The coefficients ${\rm MIG}_{2+1}^{1+1}$ correct for the migration
of $1+1$ to $2+1$ jet events from hadron to detector level
and can be determined using a fully simulated event sample
based on LEPTO. The transition from partons to hadrons is modeled
by a simple ansatz: 
$
h_{2+1}^{\rm non-pert.}(\xbx,\Qsqx)=
(\alpha + \beta \log{\xbx} + \gamma \log^2{\xbx} + \delta \log^3{\xbx})/\Qsq
$
When comparing NLO calculations to measurements, usually
the data are `corrected to parton level' by assuming that this
transition is correctly described in a LO (improved by parton showers)
Monte Carlo simulation program.
To introduce a simple hadronisation model with additional
free parameters in the fit procedure has 
the advantage that no a priori assumption
on the underlying parton level is needed to relate the
NLO calculation to the measurement.
Recently, similar models based on power behaved terms\cite{jet:powercor} 
have been very successful in describing event shapes 
in DIS\cite{jet:eventshapes}. 
In this context theoretical studies suggest that
the terms $1/Q$ may also take into account multi-emissions of soft
gluons associated with the behaviour of the running coupling constant
at small scales\cite{jet:powercor}.

In the kinematic range $25 < \Qsq < 100$\GeVsq and 
$2 \cdot 10^{-4} < \xb < 2 \cdot 10^{-2}$ jets are selected
using the $k_T$ algorithm with $d_{ij}=0.5$ and \Qsq as
scale. The jet rate for jets with $7^o < \theta_{jet}< 110^o$
and $z_q > 0.15$ is measured. LEPTO is able to describe the shape
of most distributions associated to the hard subprocess.
As an example Fig.~\ref{fig:gluon} shows the $E_T$ distribution
of the jets as measured in the detector.
LEPTO, however, fails to describe the absolute jet rate.
The measurement is always about $20$\% above the
predicted jet rate.  
This is true in the whole considered (\xbx,\Qsqx) plane,
but most pronounced for low \Qsq ($10 < \Qsq < 14$\GeVsqx) and
low \xb ($2 \cdot 10^{-4} < \xb < 1 \cdot 10^{-2}$). 
The observed difference can not be accounted for by systematic
effects. The most important systematic error of about
$3-13\%$ on the dijet rate is mainly due
to the uncertainty of the hadronic energy scale. 

The fit to the \ftwo~data and the jet rates leads to a good
$\chi^2(R_{2+1})=10$ for $11$ degrees of freedom, but no
influence of the jet data on the extracted parton density functions is
found in this analysis. Also the error band on the extracted gluon density is
only marginally improved. This means that the extracted gluon density 
is essentially constrained by the inclusive \ftwo~data. 
The fitted non-perturbative function $h(x)$ is shown in Fig.~\ref{fig:gluon}.
Without a strong
\xb dependence of $h(x)$ no satisfactory fit can be performed.

A possible explanation of this results is that the way in which
the $k_T$ jet algorithm is used in this analysis leads to
jets with relatively low transverse energy of $3-5$\GeV
(see Fig.~\ref{fig:gluon}) such that the 
found jet system does not always stem from the hard subprocess.
Another explanation is that a fixed order calculation
is not sufficient to account for the data in this phase space
region. The difference between the NLO calculation and the data
has therefore to be absorbed in the non-perturbative function $h(x)$
and consequently no sensitivity of the jet rates to the gluon density
is found.

\section{Comparision of jet rates to NLO-QCD}
A similar observation about the incapability  of NLO QCD
to describe the dijet rates has been made by the ZEUS
collaboration\cite{jet:mikunas}. In this study jets are
selected using a cone algorithm in the laboratory frame
requiring  $E_T > 4$\GeV in the hadronic center of mass
and the laboratory frame and excluding the forward region
by $\eta_{\rm lab}<2$. The data are then corrected to parton
level using LEPTO 6.3. The correction factors are 
between $1.2$ and $1.8$ for $4 \cdot 10^{-3} < \xi < 8 \cdot 10^{-2}$,
but get very large (up to $3-4$) for $\xi$ around $10^{-3}$. 
The shape of variables connected to the hard subprocess like 
$E_T$, $\eta$ and $\xi$ are well described by NLO programs. 
The absolute value of the $2+1$ cross section is, however, 
found to be about $30\%$ higher in the data.
This large difference cannot be explained by experimental effects
like variation of the energy scale or the model dependence in the correction
procedure nor by varying theoretical choices like different
parameterizations of the parton densities or different 
factorisation and renormalisation scales.
\begin{figure}
%
\epsfig{width=9cm,file=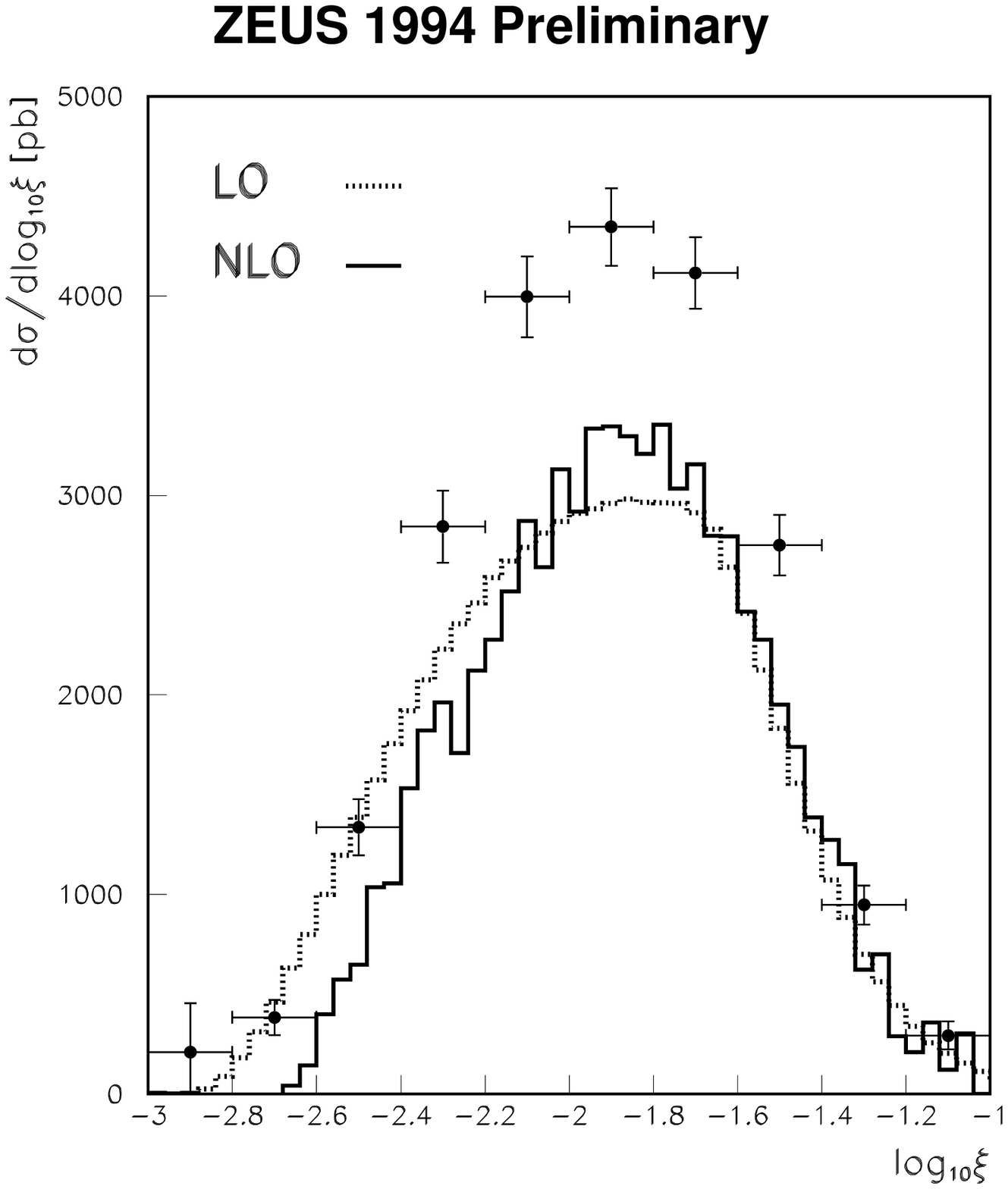}
\fcaption{ $2+1$ jet rate cross section in function of the
 momentum fraction of the parton entering the hard subprocess
 $\xi$. Data have been corrected to parton level using the
 QCD Monte Carlo program LEPTO $6.3$. Superimposed is a prediction
 from the QCD matrix element in LO and in NLO 
 in \als. 
 }
\label{fig:zeusrates}
\end{figure}

\begin{figure}
%
\epsfig{width=7cm,file=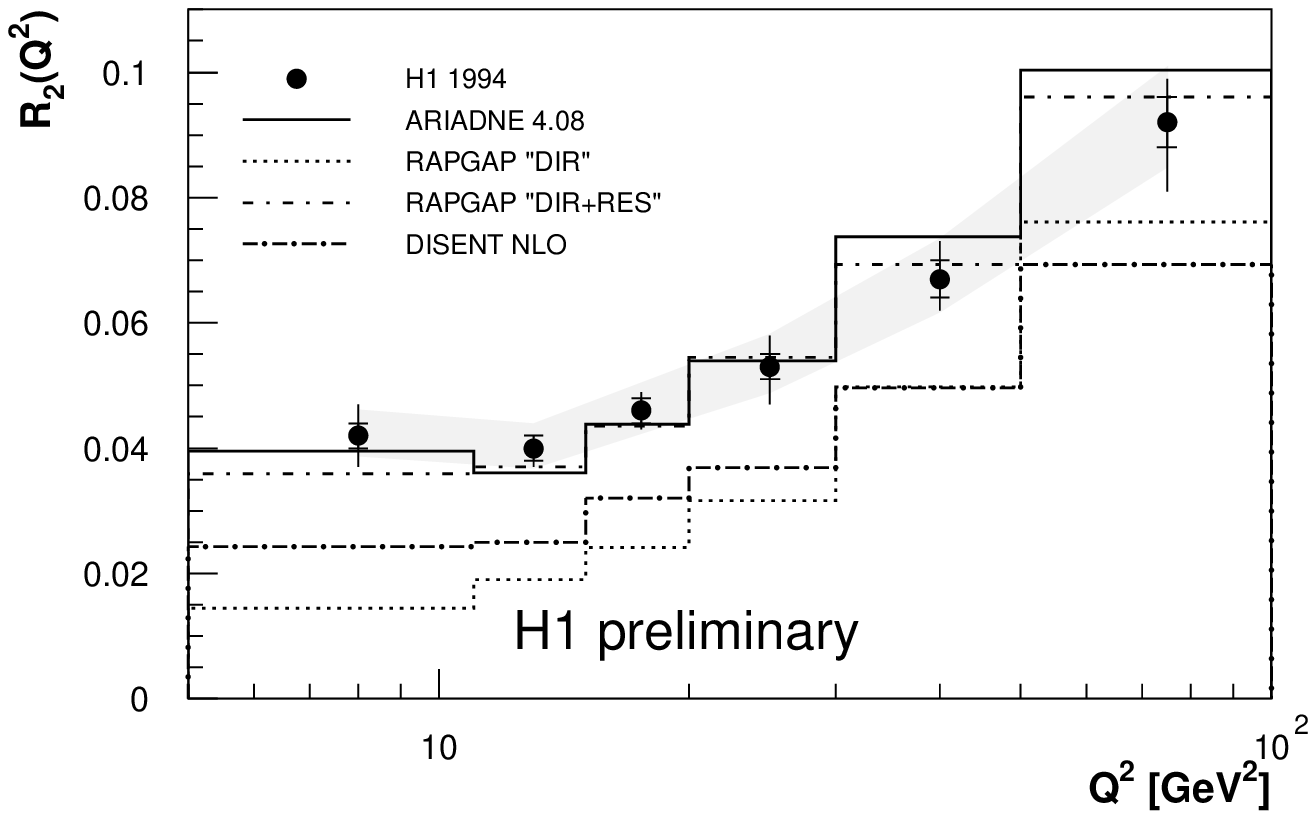}
\epsfig{width=7cm,file=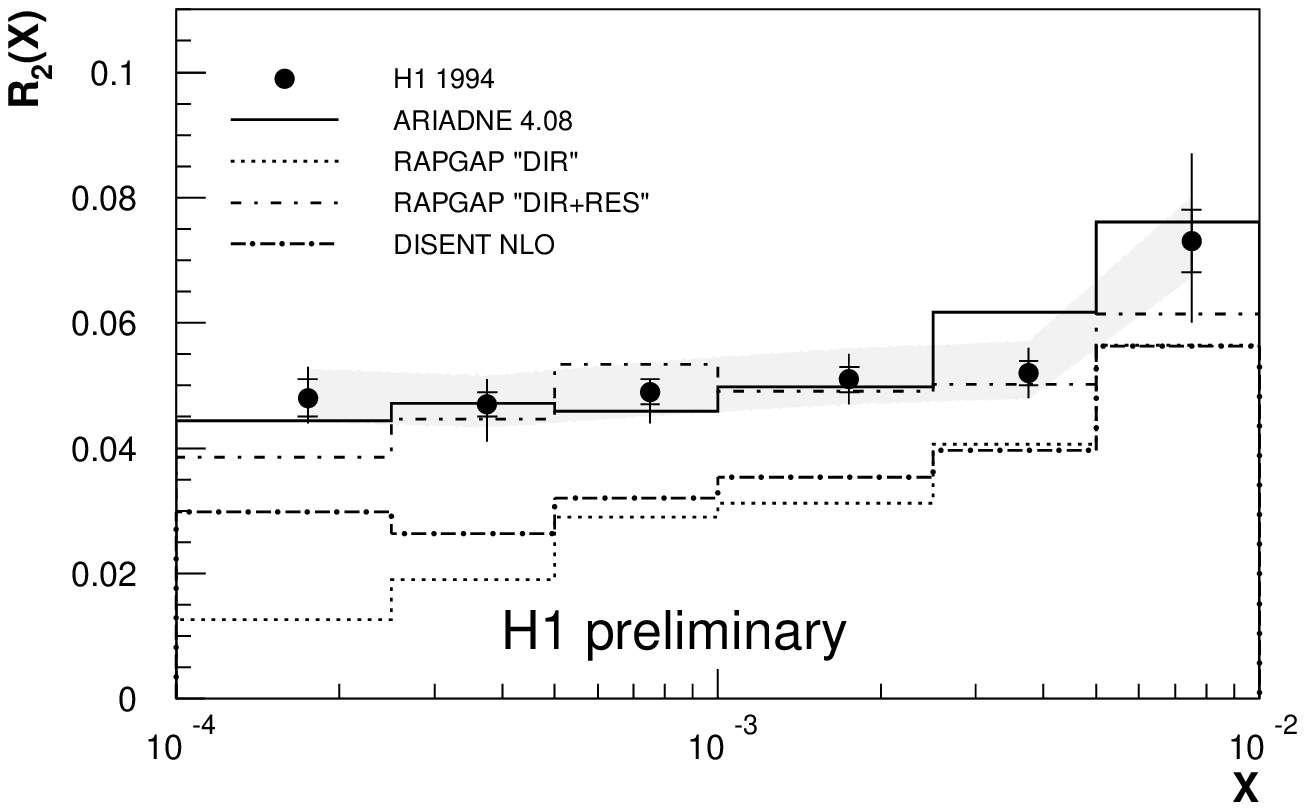}
\fcaption{$2+1$ jet rates as function of \Qsq and \xb corrected
 to hadron level. Jets are defined by the cone algorithm with
 $ E_T > 5$~\GeV in the hadronic center of mass system. 
 The difference in pseudorapidity
 between the two jets has to be smaller than $2$. The band indicates
 the systematic error. Superimposed are
 the predictions from ARIADNE, a next-to-leading calculation (DISENT) 
 and from RAPGAP (the direct and resolved contribution are shown
 separately).}
\label{fig:h1rates}
\end{figure}

That NLO QCD is not able to describe
the dijet rates has recently been reported by the 
H1 collaboration\cite{jet:spiekmor,jet:wobischdis97}.
The dijet rate is measured with a cone algorithm using a cone radius
$R=1$ and requiring $E_T > 5$\GeV in the hadronic center of mass
system. The rapidity difference between the two jets 
has to be $\Delta\eta < 2$. The $2+1$ jet rate corrected for
detector effects is shown in Fig.~\ref{fig:h1rates} as function
of \xb and \Qsqx. NLO QCD (DISENT) as well as a conventional QCD model 
based on the exact LO coefficient functions plus additional
DGLAP parton shower (RAPGAP `DIR')  clearly fail to describe the data.
This is true for the whole kinematic range 
$5 \lesssim \Qsq \lesssim 100$\GeVsq and 
$10^{-4} \lesssim \xb \lesssim 10^{-2}$, but is most pronounced
in the low \xb and \Qsq region. 
In this region the prediction of the
NLO QCD calculation is a factor of $2$ higher than the QCD model, but
is by the same factor below the data.
Varying the input parton density functions leads to a change in the
jet rate by up to $30$\%, but cannot account for the large discrepancy
with the data.

The CDM model as implemented in ARIADNE is able to describe the
data. In this model gluons are emitted from a chain of 
colour dipoles spanned between the current quark and the proton remnant.
The first emission in the cascade is corrected by the matrix element 
to first order\cite{mc:ariadneme}. It has been observed\cite{mc:rathsman} 
that quark induced processes have a up to a factor of $8$
higher cross section than expected from conventional QCD, because
the parton distribution functions in equation \ref{eq:cross} are
replaced by a phenomenological expression better adapted to the dipole picture.
Partons with short wavelengths having high transverse momenta ($P_T$) 
only resolving part of the extended colour charge distribution of the proton 
remnant are suppressed by ${(\mu/P_T)}^{\alpha}$ where
$\mu$ and $\alpha$ are phenomenological parameters usually
set to $1$.
A similar suppression depending on the virtuality \Qsq
is introduced in the direction of the scattered quark.
It has been argued that the partonic state as encountered
in ARIADNE is closely related to the one expected from a
BFKL scenario\cite{mc:bfklcdm}. In ARIADNE the hardest parton
emission is not bound to the photon vertex, but can happen
anywhere between the photon and the proton remnant.

Such a feature can also be artificially introduced in the
conventional QCD model by assuming that the photon emitted from
the incoming electron can fluctuate into quarks and gluons
before entering the hard subprocess. This can be modeled by
assuming a structure in the virtual photon like e.g. parameterized
in the SaS model\cite{mc:sasgam}.
Fig.~\ref{fig:h1rates} shows that adding this contribution
(RAPGAP `RES') to the processes where the photon is assumed to be 
point-like (RAPGAP `DIR') does also reproduce the measured jet rate.

Only models where the hardest parton emission
is not bound to occur at the photon vertex are able to describe
the data. This indicates that in the phase space region
tested by $2+1$ jet production 
higher order processes which cannot be described by NLO
QCD are important. Whether it is sufficient to simply add processes
in the photon direction without being in conflict with other HERA data, 
is not yet clear. 
The phenomenological
treatment of higher orders in the colour dipole model
has been proven to be extremely successful in describing 
all the data on the hadronic final state which have been measured at 
HERA\cite{mc:heratune,mc:carlidis96}.
It remains however a puzzle why the shape of all distributions
associated with the hard dijet system, like $E_T$, $x_p$, $\eta$ etc.,
is well described by conventional QCD. 
It is hard to imagine that the importance
of higher orders only affects the normalisation and not
the shape of the distributions.

\section{Conclusions}

The understanding of jet production in DIS remains one 
of the main physics goals at HERA.
While with the first HERA data, topics like the measurement
of the strong coupling constant and the gluon density
have been attacked with great enthusiasm, it turned
out that these goals can only be achieved in a (very) restricted
phase space region.

Existing data on jet production have always been tailored to a
specific goal and present themselves today as a loose
collection of scattered information. 
While conventional QCD is able to describe the data for
$\Qsq > 100$~\GeVsq and $\theta_{jet}>10^o$ using the JADE
algorithm, a discrepancy of about $30\%$ is found
for the low \Qsq region when using the cone or the $k_T$ algorithm
and excluding the forward region. The fully acceptance corrected jet
rates in this region disagree by about a factor of two.

It seems that before
precisely pinning down the strong coupling constant and the gluon
density, one has to systematically investigate the
applicability of NLO QCD in terms of kinematic phase
space and the way how the jet observables are defined
(jet algorithm, hard scale etc.). 

In future, it will be necessary to measure jet cross sections
corrected for detector effects
extending to very low and very large \xb and \Qsq in order
to find out where perturbative QCD is able to describe the
data and where the mechanism governing jet production are 
more complicated. To achieve this goal the understanding of 
QCD models will play a major role.

\section{Acknowledgements}
I would like to thank my colleagues J. Dainton, G. Iacobucci, M. Kuhlen,
E. Mirkes and M. Wobisch for their critical reading of the 
manuscript and helpful comments.

\section{References}
{\footnotesize

%
\end{document}